\documentclass[conference]{IEEEtran}
\IEEEoverridecommandlockouts

\usepackage{cite}
\usepackage{amsmath,amssymb,amsfonts}
\usepackage{graphicx}
\usepackage{textcomp}
\usepackage{xcolor}
\def\BibTeX{{\rm B\kern-.05em{\sc i\kern-.025em b}\kern-.08em
    T\kern-.1667em\lower.7ex\hbox{E}\kern-.125emX}}

\usepackage{amsmath, amssymb}
\usepackage{mathtools}
\usepackage[small]{caption}
\usepackage{amsthm,multirow,color,amsfonts}
\usepackage[ruled,linesnumbered]{algorithm2e}
\usepackage{tabulary}
\usepackage{graphicx}
\usepackage{setspace}
\usepackage{enumerate}
\usepackage{comment}
\usepackage{multicol,lipsum}
\usepackage{cite}
\usepackage{bm}
\usepackage{bbm}
\usepackage[noend]{algpseudocode}
\usepackage{hyperref}
\usepackage{caption}
\usepackage{subcaption}
\usepackage{textcomp}
\usepackage{xcolor}
\usepackage{subcaption}
\usepackage{booktabs,tabularx}

\makeatletter
\def\BState{\State\hskip-\ALG@thistlm}
\makeatother
	
\newtheorem{theo}{Theorem}

\newtheorem{prop}{Proposition}

\usepackage{soul}

\usepackage{ifthen}
\newif\ifFullPaper

\FullPaperfalse  

\newcommand{\showiffull}[1]{\ifFullPaper#1\fi}  
\newcommand{\showifshort}[1]{\ifFullPaper\else#1\fi}

\IEEEaftertitletext{\vspace{-1\baselineskip}}
    
\begin{document}

\title{Decentralized AI Service Placement, Selection and Routing in
Mobile Networks
}

\author{\IEEEauthorblockN{Jinkun Zhang\quad Stefan Vlaski \quad Kin Leung}
\IEEEauthorblockA{
Imperial College London, UK}
}

\maketitle

\begin{abstract}
The rapid development and usage of large-scale AI models by mobile users will dominate the traffic load in future communication networks. 
The advent of AI technology also facilitates a decentralized AI ecosystem where small organizations or even individuals can host AI services. 
In such scenarios, AI service (models) placement, selection, and request routing decisions are tightly coupled, posing a challenging yet fundamental tradeoff between service quality and service latency, especially when considering user mobility.
Existing solutions for related problems in mobile edge computing (MEC) and data-intensive networks fall short due to restrictive assumptions about network structure or user mobility. 
To bridge this gap, we propose a decentralized framework that jointly optimizes AI service placement, selection, and request routing. 
In the proposed framework, we use \emph{traffic tunneling} to support user mobility without costly AI service migrations. 
To consider nonlinear queueing delays, we formulate a non-convex problem to optimize the tradeoff between service quality and the end-to-end latency.
We derive the node-level KKT conditions and develop a decentralized Frank–Wolfe algorithm with a novel messaging protocol. 
Numerical evaluations are used to validate the proposed approach and show substantial performance improvements over existing methods.
\end{abstract}

\section{Introduction}
The rapid adoption of AI services (e.g., OpenAI's GPT series) is fundamentally changing the traffic load and dynamics of modern communication networks. 
While AI services today are primarily offered by major companies,  predictions (e.g., \cite{forbes2024democratizingai}) point towards a more decentralized future AI ecosystem, where small organizations or even individual users can host their own AI models, presumably in decentralized networks with flexible scales and arbitrary topologies.

This poses significant challenges for both users and networks.
Users have the options of selecting from multiple pre-trained AI models offered by different providers. 
These models provide different levels of service quality (e.g., accuracy) and latency, requiring users to carefully select the one that best aligns with their preferences \cite{ hadjkouider2024review}.
The network, on the other hand, should carefully place the models to keep network congestion and user latency under control.
{Recent studies on \emph{AI as a network service} examine model placement and resource optimization under latency/accuracy goals~\cite{ huang2024ai}, and selection across models with heterogeneous QoS~\cite{hudson2024qos}, but most assume centralized control or limited topologies and do not target fully decentralized settings.}


\begin{figure}[htbp]
\centerline{\includegraphics[width=0.9\linewidth]{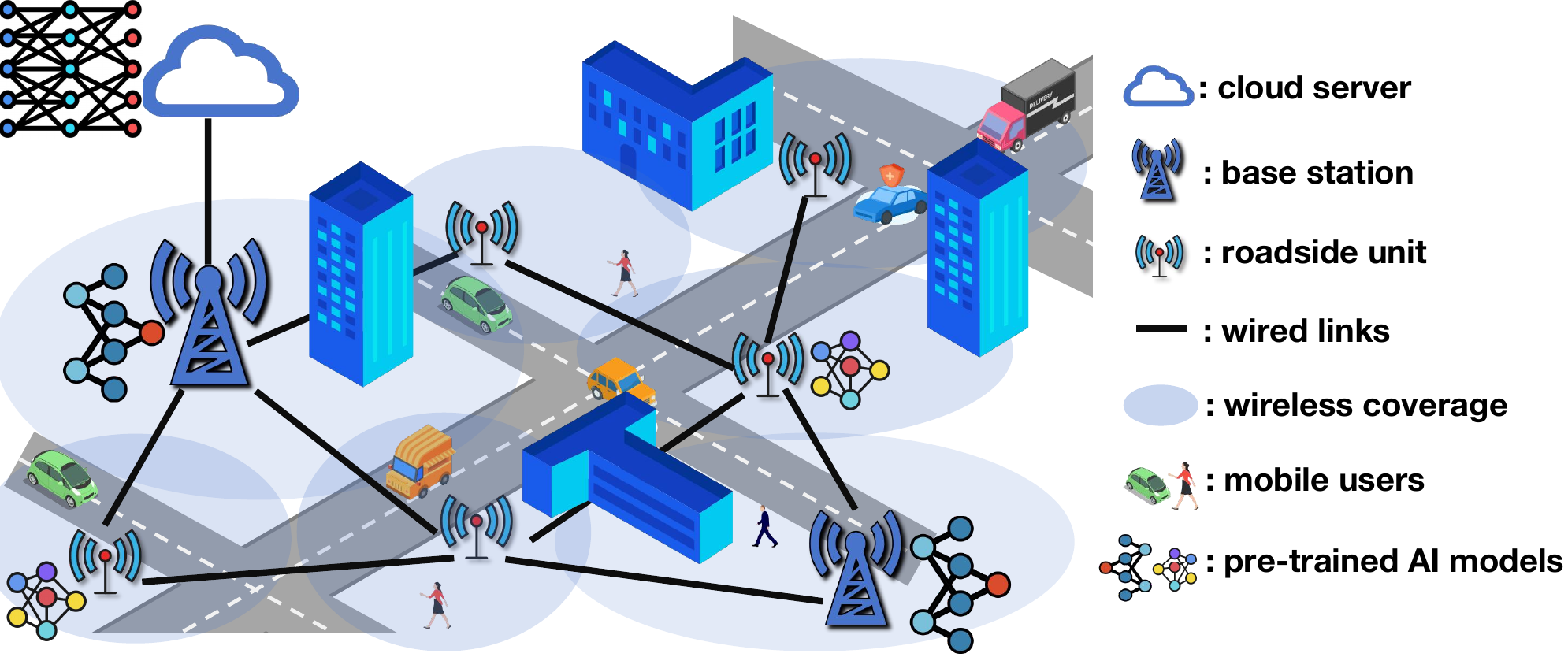}}
\vspace{-0.2\baselineskip}
\caption{An example edge-cloud vehicular network. Mobile users have multiple pre-train AI model options.}
\label{fig_network}
\vspace{-0.8\baselineskip}
\end{figure}

First, AI service placement and selection are very similar to those in mobile edge computing (MEC).
In MEC, service selection focuses on choosing the most appropriate service instance to balance performance and efficiency under constraints such as latency, quality of service (QoS), and hardware limitations \cite{wu2019mobility}.
Service placement aims to provide popular services or network content close to users to reduce access delay and network load \cite{ li2023optimal}.
{However, these MEC approaches generally rely on hierarchical control and are not designed for decentralized scenarios.}
Although another line of studies for content and computation placement support arbitrary decentralized topologies \cite{cai2023joint, zhang2024congestion}, they assume pure static networks and overlook user mobility.

On the other hand, unlike traditional MEC, AI models introduce additional challenges. Most notably, the size of pre-trained AI models is growing exponentially (e.g., GPT-4 has roughly 1.8 trillion parameters), making it increasingly difficult for edge servers to host. In conventional MEC systems, when users move between wireless access points (APs), \emph{service migration} is commonly employed to maintain seamless service delivery \cite{wang2019dynamic,liang2021multi}. It transfers the service instance and its runtime state to a server closer to the user's new location, assuming the service is lightweight enough to be moved.
However, the cost of real-time migrating large AI models is considered impractical \cite{tu2025distributed}.
To address this, we adopt a more classic and realistic solution: \emph{traffic tunneling} \cite{taleb2017multi}, where the user’s original AP serves as an anchor. 
When the user moves, responses from the remote server are first routed back to this anchor node, which then forwards the results to the user’s new location. This approach eliminates the need to migrate large-scale AI models, instead incurring overhead in the form of additional traffic flows.
A positive feedback loop exists between request latency and tunneling flows, which may overload the network if not appropriately handled.  
To our knowledge, traffic tunneling-based AI service placement and selection has never been studied.

In this work, we address the above gaps by developing a novel framework that jointly optimizes AI service placement, selection, and request routing under user mobility. The proposed framework supports arbitrary network topologies and operates fully decentralized via traffic tunneling. 
A key contribution of our approach is the use of congestion-dependent nonlinear costs, which can capture the crucial queueing effects on network links and processing units. 
Rather than modeling individual requests, we analyze the time-averaged system behavior under homogeneous assumptions.
We tackle the non-convex optimization problem first with fixed service placement and then extend to the general case. 
We derive node-level Karush--Kuhn--Tucker (KKT) conditions, which indicate intuitively myopic node behavior. 
We then give decentralized online Frank--Wolfe-based algorithms that converge to these conditions via a novel messaging protocol to obtain gradients.

Our major contributions are summarized as follows:
\begin{itemize}
    \item We propose a decentralized framework that jointly handles AI service placement, selection, and routing under user mobility using a tunneling mechanism.
    \item We formulate a utility-minus-cost optimization problem with congestion-aware costs, and derive node-level KKT conditions for both fixed placement and the general case.
    \item We design a decentralized messaging protocol and algorithms convergent to the KKT conditions, then validate their performance through numerical evaluations.
\end{itemize}

The paper first tackle the problem under fixed service placement in Section~\ref{sec:fixplacement} and \ref{sec:decentralized_alg}, then extend to joint service placement in Section~\ref{sec:joint_placement} and provide numerical evaluation results in Section~\ref{sec:simulation}.
\showifshort{Due to space limit, we put the proofs of propositions and theorems in supplementary document \cite{supplementary2025}.}

\section{Fixed AI Service Placement}
\label{sec:fixplacement}

\subsection{AI-driven network with mobile users}
\noindent\textbf{Network and AI services.}
We consider a directed, connected graph $\mathcal{G} = (\mathcal{V}, \mathcal{E})$ with arbitrary topology, where $\mathcal{V}$ are static nodes (e.g., APs, RSUs, edge servers), and $\mathcal{E}$ are links.
For node $i$, let $\mathcal{N}_i$ denote $i$'s neighbors $\{j:(i,j)\in\mathcal{E}\}$.
Nodes in $\mathcal{V}$ have heterogeneous capabilities of communication with neighbors, hosting pre-trained AI models, and serving inference requests.
Mobile users access the network by associating with nodes. Let $\mathcal{U}$ be the set of users, and assume user $u \in \mathcal{U}$ is associated with node $v_u(t)$ at time slot $t$. 
Communication and computation in the network are driven by a set of AI tasks $\mathcal{K}$.
Every task $k \in \mathcal{K}$ can be fulfilled by a set of (different) pre-trained models $\mathcal{M}_k$ in the network.
We define a \emph{service} as a pair $(k,m)$ for $k\in\mathcal{K}$ and $m\in\mathcal{M}_k${\footnote{ $m=0$ for lightweight local models  (e.g., onboard obstacle detection).}}, and $\mathcal{S}$ be the set of services.
In this section, we assume fixed service placement: each $(k,m)$ is hosted by a known, non-empty set $\mathcal{X}_{k,m} \subseteq \mathcal{V}$.

\vspace{0.3\baselineskip}
\noindent\textbf{Requests handling.}
{\color{red}
}
During time slot $t$, user $u$ issues a request for task $k$ with probability $r_u^{k}(t) \in [0,1]$.
For task $k$, user $u$ can specify models in $\mathcal{M}_k$ by \emph{model selection decision} $s_{u}^{k,m}(t)\in [0,1]$, i.e., a fraction of $s_{u}^{k,m}(t) $ of rate $r_{u}^k(t)$ is assigned to model $m \in \mathcal{M}_k$ with $\sum\nolimits_{m \in \mathcal{M}_k} s_{u}^{k,m}(t) = 1.$
After model selection, requests $(k,m)$ for $m \neq 0$ are routed through the network in a hop-by-hop distributed manner to a node that hosts the service (i.e., in set $\mathcal{X}_{k,m}$).
If the user remains stationary, the inference result is then delivered back along the reversed path of the request (we defer the mobility handling right after).
The network routing is controlled by routing decisions $\phi_{ij}^{k,m}(t)$. 
Specifically, requests for $(k,m)$ arrive at node $i$ from two sources:
(i) exogenous requests, issued by users currently associated with $i$, and
(ii) endogenous requests, forwarded from neighboring nodes. 
Of all arrival $(k,m)$-requests, node $i$ forwards a fraction of $\phi_{ij}^{k,m}(t) \in [0,1]$ to each neighbor $j \in \mathcal{N}_i$, with the flow conservation
$\sum\nolimits_{j\in\mathcal{N}_i}\phi_{ij}^{k,m}(t) = \mathbbm{1}_{i \not\in \mathcal{X}_{k,m}}$. 
Namely, if $i \in \mathcal{X}_{k,m}$, it provide service $(k,m)$ and act as a sink of requests; otherwise, it forwards all arrival requests to neighbors. 
Such a hop-by-hop routing scheme aligns naturally with decentralized control and has been widely adopted in data- and computation-intensive networks \cite{gallager1977minimum,zhang2024congestion}.
Conceptually similar to MoE gating~\cite{li2025theory}, we consider choice-dependent utility, capturing heterogeneous AI outputs:  for each fulfilled request of service $(k,m)$, the user obtains a utility $u_{k,m} \geq 0$, reflecting the inference quality (e.g., accuracy or user satisfaction) of the selected model.

\begin{figure}[t]
  \centering
  \begin{subfigure}[t]{0.52\columnwidth}
    \centering
    \includegraphics[width=\linewidth]{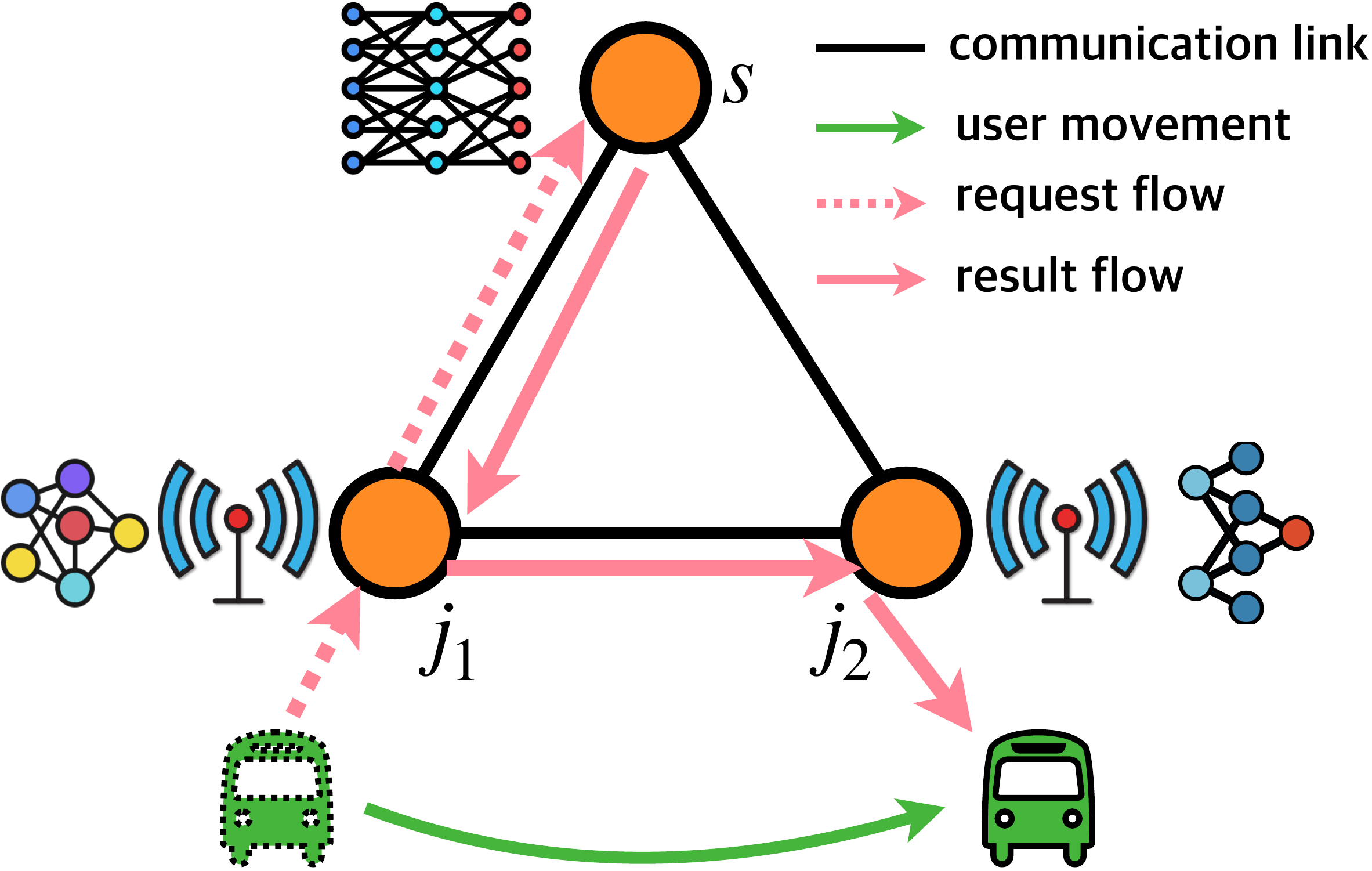}
    \caption{Tunneling on $(j_1,j_2)$}
    \label{fig:tunneling1}
  \end{subfigure}
  \hfill
  \begin{subfigure}[t]{0.45\columnwidth}
    \centering
    \includegraphics[width=\linewidth]{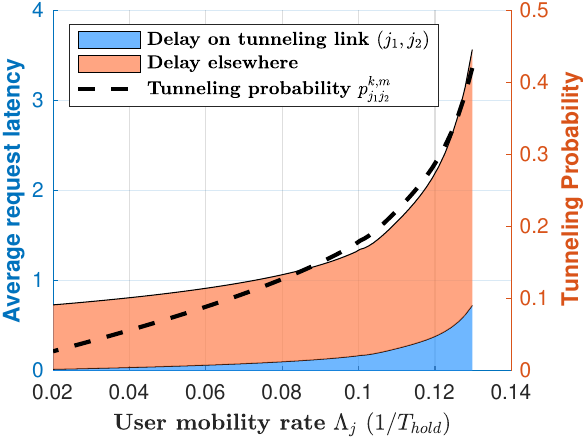}
    \caption{Latency $D_{j_1,k,m}$ and $p_{j_1j_2}^{k,m}$}
    \label{fig:tunneling2}
  \end{subfigure}
  \vspace{-0.5\baselineskip}
  \caption{Traffic tunneling and impact on request latency.}
  \label{fig:tunneling_illustration}
  \vspace{-1\baselineskip}
\end{figure}

\vspace{0.3\baselineskip}
\noindent\textbf{Traffic tunneling.}
Since large AI models are impractical to migrate in real time, we adopt \emph{traffic tunneling} to handle user mobility.
Suppose user $u$ issues a request for service $(k,m)$ at time $t$, the node $ i = v_u(t)$ serves as an ``anchor'', so that the result is always sent back to $i$ upon completion.
Let $D^{\text{o}}_{i,k,m}$ denote the ``static'' round-trip latency of the request, measured from it being generated until the result returns to $i$.
If the user has moved during this time, i.e., $v_u(t + D^{\text{o}}_{i,k,m}) = j \neq i$, then the anchor node $i$ will forward the result to the new access point $j$ to reach the user.
We assume tunneling occurs over at most one hop, i.e., $v_u(t + D^{\text{o}}_{i,k,m}) \in \mathcal{N}_{v_u(t)}$.{\footnote{ Multi-hop transitions are extremely rare in practice.
}}
{Such tunneling scheme (Figure~\ref{fig:tunneling_illustration}(a)) aligns with our decentralized setting,} 
however, inducing a positive feedback loop: tunneling adds extra transmission, increasing overall latency, which in turn triggers more tunneling.
We aim to manage the tunneling flow to stabilize and optimize the system.

\subsection{Time-homogeneous formulation}
\label{sec:quasi-static}
\noindent\textbf{Time-homogeneous network.}
We adopt time-homogeneous approximations to simplify the system dynamics. Assume the aggregated request rates received at each AP are quasi-static, and the request for $k$ received at node $i$ is time-invariant $r_{i}^k$, 
\begin{equation}
   \sum\nolimits_{u \in \mathcal{U}: v_u(t) = i} r_{u}^k(t) = r_{i}^k, \quad \forall t.
\end{equation}

We also replace the time- and user-dependent service selection $s_{u}^{k,m}(t)$ with a node-based time-invariant variable $s_{i}^{k,m}$,
\begin{equation}
    s_{u}^{k,m}(t) = s_{i}^{k,m}, \quad \forall u:v_u(t) = i, \,\,\forall t.
\end{equation}
We denote by vector $\boldsymbol{s} = [s_{i}^{k,m}]$ the \emph{global service selection variable}, with the following constraint holds,
\begin{equation}
    \sum\nolimits_{m \in \mathcal{M}_k} s_{i}^{k,m} = 1, \quad \forall i \in \mathcal{V}, k \in \mathcal{K}.
    \label{constraint_s}
\end{equation}

Similarly, we use time-invariant routing variable $\phi_{ij}^{k,m}$ with 
\begin{equation}
    \phi_{ij}^{k,m} = \phi_{ij}^{k,m}(t), \quad\forall t,
\end{equation}
and denote by vector $\boldsymbol{\phi} = [\phi_{ij}^{k,m}]$ the \emph{global routing variable} with the flow conservation given by
\begin{equation}
    \sum\nolimits_{j\in\mathcal{N}_i}\phi_{ij}^{k,m} = \mathbbm{1}_{i \not\in \mathcal{X}_{k,m}}, \quad\forall i \in\mathcal{V}, (k,m) \in \mathcal{S}.
    \label{constraint_phi}
\end{equation}
We remark that in practice, $\boldsymbol{\phi}$ can be implemented by simple probabilistic request forwarding, i.e., when node $i$ receives a request for $(k,m)$, it forwards it to $j$ with probability $\phi_{ij}^{k,m}$.

\vspace{0.3\baselineskip}
\noindent\textbf{Service latency.}
Request latency is incurred at both traversed links and the service-hosting node.
We assume both delays are dependent on the link flow or node workload, capturing the crucial queueing effect.
Let $f_{ij}^{k,m}$ be the steady-state request rate of service $(k,m)$ on link $(i,j) \in \mathcal{E}$, given by
\begin{equation}
    f_{ij}^{k,m} = \phi_{ij}^{k,m} t_i^{k,m},
\end{equation}
where $t_i^{k,m}$ is the total received request rate for $(k,m)$ at node $i$, given recursively by
\begin{equation}
    t_i^{k,m} = r_{i}^k s_{i}^{k,m} + \sum\nolimits_{j \in \mathcal{N}_i} f_{ji}^{k,m}.    
\end{equation}

Let $L^{\text{req}}_{k,m}$ and $L^{\text{res}}_{k,m}$ be the size of request and result packets, respectively, then the total flow rate on $(i,j)$ is
\begin{equation}
    F_{ij} = F_{ij}^{\text{o}} + F_{ij}^{\text{tun}},
\end{equation}
where $F_{ij}^{\text{o}}$ is the \emph{static flow} given by
\begin{equation}
F_{ij}^{\text{o}} = \sum\nolimits_{(k,m)\in\mathcal{S}}\left(L^{\text{req}}_{k,m}f_{ij}^{k,m} + L^{\text{res}}_{k,m}f_{ji}^{k,m}  \right),
\end{equation}
and $F_{ij}^{\text{tun}}$ is the \emph{tunneling flow} (extra flow due to traffic tunneling), given later in \eqref{F_tun}. 
We denote the expected packet delay on link $(i,j)$ by $d_{ij}(F_{ij})$, where $d_{ij}(\cdot)$ is non-decreasing and convex. e.g., the average M/M/1 queue sojourn time~\cite{bertsekas2021data}
\begin{equation}
d_{ij} = 1/({\mu_{ij} - F_{ij}}),
\end{equation}
where $\mu_{ij}$ is the service rate (i.e., capacity), given $F_{ij} < \mu_{ij}$.

Let $W_{k,m}$ be the single-request computation workload for $(k,m)$, the total workload at node $i$ is
\begin{equation}
    G_i = \sum\nolimits_{(k,m):i \in \mathcal{X}_{k,m}} W_{k,m}t_i^{k,m}, \quad \forall i \in \mathcal{V}
\end{equation}
Similarly, let $c_i(G_i)$ be the expected request delay at node $i$, where $c_i(\cdot)$ is non-decreasing convex, incorporating delay due to computation processing and congestion effect.

The end-to-end latency of a request comprises three parts: 
(i) request transmission delay, 
(ii) delay at the service-hosting node, and
(iii) result transmission delay (on the reversed path). 
Thus, the static round-trip delay $D^{\text{o}}_{i,k,m}$ is given by
\begin{equation*} 
\sum\nolimits_{p \in \mathcal{P}_i^{k,m}} \mathbb{P}_p\left( c_{p_{|p|}} +  \sum\nolimits_{\ell=1}^{|p|-1} \left( d_{p_\ell p_{\ell+1}} + d_{p_{\ell+1} p_\ell} \right) \right) + d_{\text{AP}}, 
\end{equation*} 
where $p$ is a \emph{routing path}, i.e., node sequence $(p_1, p_2,\cdots,p_{|p|})$ with $\phi_{p_l p_{l+1}} > 0$; set $\mathcal{P}_i^{k,m}$ denotes all paths for service $(k,m)$ starts from $i$, i.e., $p_1 = i$, $p_{|p|}\in\mathcal{X}_{k,m}$;
constant $d_{\text{AP}}$ is the user-AP wireless access delay;
 and $\mathbb{P}_p = \prod_{l=1}^{|p|}\phi_{p_l p_{l+1}}$ is the probability that path $p$ is taken under $\boldsymbol{\phi}$.

Moreover, the expected end-to-end latency of service $(k,m)$ issued originally at $i$ is: 
\begin{equation} 
D_{i,k,m} = D^{\text{o}}_{i,k,m} + \sum\nolimits_{j \in \mathcal{N}_i} p_{ij}^{k,m} d_{ij},
\label{D_ikm}
\end{equation}
where $p_{ij}^{k,m}$ is the \emph{tunneling probability}, i.e., the chance that user moved to $j$ during time $D^{\text{o}}_{i,k,m}$,  given later in \eqref{p_ij}.
\footnote{For local models $m = 0$, we assume $D^{i,k,0} = c_u$ with a constant $c_u$.}

\vspace{0.3\baselineskip}
\noindent\textbf{Tunneling flow.}
We assume users have homogeneous movement patterns with the classic \emph{continuous-time Markov chain} model~\cite{jalel2020continuous}.
Let $\lambda_{ij} \geq 0$ be user transition rate from $i$ to $j$, and let $\Lambda_{i} = \sum_{j\in\mathcal{N}_i} \lambda_{ij}$, the user association time at $i$ denoted by $T^{\text{hold}}_{i}$ follows an exponential distribution 
\begin{equation}
f_{T^{\text{hold}}_{i}}(T) = \Lambda_{i} e^{-\Lambda_{i} T}, \quad T \geq 0.
\end{equation}
After association ends, the probability of transition to $j$ is:
\begin{equation}
    q_{ij} = \lambda_{ij}/\Lambda_{i}.
\end{equation}
Therefore, the tunneling probability during $D^{\text{o}}_{i,k,m}$ is
\begin{equation}
    p_{ij}^{k,m} = q_{ij}\mathbb{P}\{ D^{\text{o}}_{i,k,m} > T^{\text{hold}}_{i}\} = q_{ij}\left(1 - e^{-\Lambda_{i}D^{\text{o}}_{i,k,m}}\right)
    \label{p_ij}
\end{equation}
and recall that tunneling flow on $(i,j)$ is incurred by $i$ forwarding result packets of users moved from $i$ to $j$, we have
\begin{equation}
F^{\text{tun}}_{ij} = \sum\nolimits_{(k,m) \in \mathcal{S}} L^{\text{res}}_{k,m} r_i^{k} s_i^{k,m} p_{ij}^{k,m}.
\label{F_tun}
\end{equation}
{Figure~\ref{fig:tunneling_illustration}(b) illustrates that tunneling can significantly increase overall service latency as user mobility intensifies.}


\vspace{0.3\baselineskip}
\noindent\textbf{Quality-latency tradeoff.}
Let $\eta$ be the system’s quality-latency tradeoff preference, we maximize the average request utility minus latency over service selection and routing:
\begin{align}
    \max_{\boldsymbol{s}, \boldsymbol{\phi} } \,\, &Q=\frac{\sum_{i\in\mathcal{V}}\sum_{k\in\mathcal{K}}r_i^k \sum_{m\in\mathcal{M}_k}s_i^{k,m}\left(\eta u_{k,m} - D_{i,k,m}\right)}{\sum_{i\in\mathcal{V}} \sum_{k\in\mathcal{K}}r_i^k} \nonumber
\\ \text{s.t.} \,\,\, &  \boldsymbol{s} \geq \boldsymbol{0}, \, \boldsymbol{\phi} \geq \boldsymbol{0}, \, \text{ and \eqref{constraint_s},\eqref{constraint_phi} hold} 
    \tag{P0}
\label{obj_origin}
\end{align}

Problem \eqref{obj_origin} captures the tradeoff between AI service quality and experienced latency across all requests. 
It is difficult as $D_{i,k,m}$ is highly coupled through links along the paths. We reformulate \eqref{obj_origin} into a more tractable form. 
\begin{align}
\min_{\boldsymbol{s},\boldsymbol{\phi}} \,\, &J=\sum_{(i,j)\in\mathcal{E}} D_{ij} + \sum_{i\in\mathcal{V}\cup\mathcal{U}} C_i -  \sum_{i\in\mathcal{V}}\sum_{(k,m)\in\mathcal{S}}
    \hat{u}_{k,m}r_i^ks_i^{k,m} \nonumber
    \\ \text{s.t.} \,\,\, &  \boldsymbol{s} \geq \boldsymbol{0}, \, \boldsymbol{\phi} \geq \boldsymbol{0}, \, \text{ and \eqref{constraint_s},\eqref{constraint_phi} hold}
    \tag{P1}
\label{obj_time_invar}
\end{align} 

where $D_{ij} = F_{ij} d_{ij}$, $C_i = G_i c_i$, and $\hat{u}_{k,m} = \eta u_{k,m} - d_{AP}\mathbbm{1}_{m \neq 0}$ is the modified utility.

\begin{prop}
    With fixed $\{r_i^k\}$,  \eqref{obj_time_invar} is equivalent to \eqref{obj_origin}. Specifically, it holds $J = -\left(\sum_{i}\sum_kr_i^k\right)Q$ for any $(\boldsymbol{s},\boldsymbol{\phi})$.
    \label{prop_homo}
\end{prop}

We remark that $\eqref{obj_time_invar}$ is non-convex in $(\boldsymbol{s},\boldsymbol{\phi})$. While~\eqref{obj_time_invar} structurally resembles previous models (e.g., \cite{zhang2024loam}), the inclusion of tunneling introduces significant complexity. 
Theorem \ref{thm_KKT_no_cache} gives a set of KKT necessary optimality conditions for \eqref{obj_time_invar}.
It aligns naturally with decentralized decision-making, and nodes need only act ``myopically'' based on the marginal costs. 
e.g., for service selection, user at $i$ should assign new requests to the minimum-marginal model $\partial J/\partial s_i^{k,m}$; for routing, node $i$ should forward marginal incoming requests of $(k,m)$ to the neighbor $j$ minimizing $\partial J/\partial \phi_{ij}^{k,m}$. 

\begin{theo}
    Suppose $(\boldsymbol{\phi},\boldsymbol{s})$ optimally solves \eqref{obj_time_invar}, then
\begin{subequations}
\begin{gather}
    \frac{\partial J}{\partial s_{i}^{k,m}} \begin{cases}
        = \min_{n \in \mathcal{M}_k} \frac{\partial J}{\partial s_{i}^{k,n}}, \quad \text{if } s_{i}^{k,m} > 0,
        \\ \geq \min_{n \in \mathcal{M}_k} \frac{\partial J}{\partial s_{i}^{k,n}}, \quad \text{if } s_{i}^{k,m} = 0,
    \end{cases} \label{KKT_no_cache_s}
    \\
     \frac{\partial J}{\partial \phi_{ij}^{k,m}} \begin{cases}
        = \min_{l \in \mathcal{N}_i} \frac{\partial J}{\partial \phi_{il}^{k,m}}, \quad \text{if } \phi_{ij}^{k,m} > 0,
        \\ \geq \min_{l \in \mathcal{N}_i} \frac{\partial J}{\partial \phi_{il}^{k,m}}, \quad \text{if } \phi_{ij}^{k,m} = 0.\label{KKT_no_cache_phi}
    \end{cases}
\end{gather}
\label{KKT_no_cache}
\end{subequations}
\label{thm_KKT_no_cache}
\end{theo}
\showiffull{
\begin{proof}
    Theorem \ref{thm_KKT_no_cache} is a special case of Theorem \ref{thm_KKT_cache}, which will be introduced in Section \ref{sec:joint_placement} and proved in Appendix \ref{proof_KKT_cache}.
\end{proof}
}
In general, condition \eqref{KKT_no_cache} only guarantees the necessity for optimality.
Nevertheless, for a simplified system with linear (congestion-independent) costs, it is sufficient for optimality.

\begin{prop}
    Suppose $r_i^{k} > 0$ for all $i\in\mathcal{V}$ and $k \in \mathcal{K}$. 
    If $d_{ij}(F_{ij}) = d_{ij}$ with constant $d_{ij} > 0$ for all $(i,j) \in \mathcal{E}$, and $c_i(G_i) = c_i$ with constant $c_i > 0$ for all $i \in \mathcal{V \cup \mathcal{U}}$, then any $(\boldsymbol{s},\boldsymbol{\phi})$ feasible to \eqref{obj_time_invar} and satisfying \eqref{KKT_no_cache} is a global optimizer.
\label{prop_sufficient}
\end{prop}

\section{Decentralized Algorithm Design}
\label{sec:decentralized_alg}
In this section, we propose a decentralized method to obtain $(\boldsymbol{s},\boldsymbol{\phi})$ that satisfies KKT condition \eqref{KKT_no_cache}.
We first decompose gradients $\partial J / \partial s_i^{k,m}$ and $\partial J / \partial\phi_{ij}^{k,m}$ involved in \eqref{KKT_no_cache}. Let
\begin{equation}
    \tilde{t}_i^{k,m} = \sum\nolimits_{j \in \mathcal{N}_i} f_{ji}^{k,m}
\end{equation}
denote the endogenous arrival rate for $(k,m)$ at $i$, then $t_i^{k,m} = r_i^ks_i^{k,m} + \tilde{t}_i^{k,m}$.
For $m \neq 0$, let $\delta_{i}^{k,m}$ be the marginal latency caused by increased $\tilde{t}_i^{k,m}$, and
$\tau_i^{k,m}$ be the marginal latency caused by tunneling increased exogenous arrival rate $r_i^ks_i^{k,m}$,
\begin{align}
    \delta_{i}^{k,m} &= {\partial \left(\sum\nolimits_{(p,q)\in\mathcal{E}}D_{pq} + \sum\nolimits_{p\in\mathcal{V}\cup\mathcal{U}}C_p\right)}\big/{\partial \tilde{t}_i^{k,m}}, \label{delta}
    \\ \tau_i^{k,m} &= L_{k,m}^{\text{res}}\sum\nolimits_{j \in \mathcal{N}_i} D^\prime_{ij}(F_{ij}) p_{ij}^{k,m}. \label{tau}
\end{align}
Then the gradients can be decomposed using $\delta_i^{k,m}$ and $\tau_i^{k,m}$.

\begin{figure}[htbp]
\centerline{\includegraphics[width=1\linewidth]{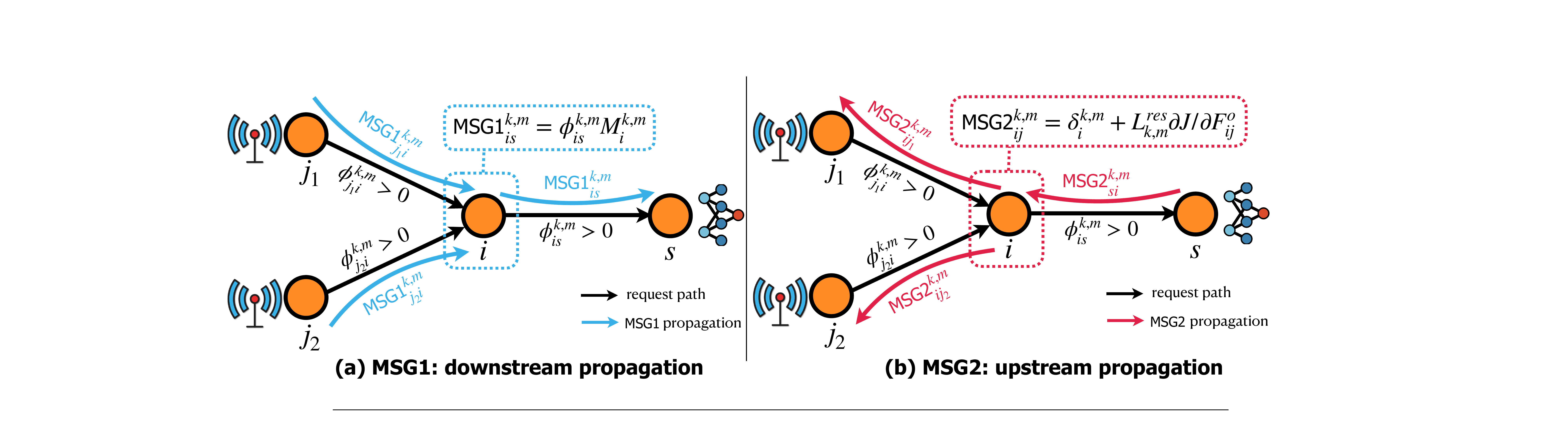}}
\vspace{-0.3\baselineskip}
\caption{Illustration of $\texttt{MSG1}$ and $\texttt{MSG2}$ propagation in DMP.}
\vspace{-0.3\baselineskip}
\label{fig:DMP}
\end{figure}

\begin{theo}
\label{theo_recursive}
For $m = 0$ (local models), 
\begin{subequations}
\begin{equation}
    {\partial J}/{\partial s_i^{k,0}} = r_i^k (W_{k,m}c_u - \hat{u}_{k,m}).
    \label{pJps_local}
\end{equation}

For $m \neq 0$, 
\begin{equation}
    {\partial J}/{\partial s_{i}^{k,m}} = r_{i}^k \left(\delta_i^{k,m} + \tau_i^{k,m} - \hat{u}_{k,m}\right),
    \label{pJps}
\end{equation}
\begin{equation}
\frac{\partial J}{\partial \phi_{ij}^{k,m}} = t_{i}^{k,m}\big(L_{k,m}^{\text{req}}\frac{\partial J}{\partial F_{ij}^{\text{o}}}+ L_{k,m}^{\text{res}}\frac{\partial J}{\partial F_{ji}^{\text{o}}}+\delta_{j}^{k,m}\big).
\label{pJpphi}
\end{equation}
\label{partial_J}
\end{subequations}

For $i \in \mathcal{X}_{k,m}$, 
\begin{subequations}
\begin{equation}
\delta_{i}^{k,m} = W_{k,m}C^\prime_i(G_i).
\label{delta_cache}
\end{equation}

For $i \not\in \mathcal{X}_{k,m}$, $\delta_{i}^{k,m}$ is recursively given by
\begin{equation}
    \delta_{i}^{k,m} = \sum_{j \in \mathcal{N}_i}\phi_{ij}^{k,m}\big(L_{k,m}^{\text{req}}\frac{\partial J}{\partial F_{ij}^{\text{o}}}+ L_{k,m}^{\text{res}}\frac{\partial J}{\partial F_{ji}^{\text{o}}}+\delta_{j}^{k,m}\big).
    \label{delta_no_cache}
\end{equation}
\label{delta_recursive}
\end{subequations}
\label{thm_decentralize}
\end{theo}
We will give $\partial J / \partial F_{ij}^{\text{o}}$ in \eqref{pJpFo}.
Theorem~\ref{thm_decentralize} is the first to generalize the classic recursive gradient decomposition in \cite{gallager1977minimum} to analytically incorporate user mobility. 
When users are static $\lambda_{ij} = 0$ and request/result sizes simplify to $L_{k,m}^{\text{req}}=1$, $L_{k,m}^{\text{res}}=0$, the above recovers the analysis in~\cite{gallager1977minimum} exactly.
Based on Theorem~\ref{thm_decentralize}, using only local and neighbor information, we design a Decentralized Messaging Protocol (DMP) to estimate $\partial J/\partial s_i^{k,m}$ and $\partial J / \partial \phi_{ij}^{k,m}$.
We provide general ideas of DMP and refer readers to \cite{supplementary2025} for details. Figure~\ref{fig:DMP} illustrates DMP at a relay node $i$ connected with APs $j_1$, $j_2$ and server $s$.

In DMP, two types of control messages, $\texttt{MSG1}$ and $\texttt{MSG2}$, are propagated in the network. 
The intuition of DMP builds on classic recursive messaging schemes \cite{zhang2024congestion}.
In \cite{zhang2024congestion}, control messages propagate upstream along request paths, to inform each node of downstream network status and thus adjust local routing decisions. In our case, this idea is retained in $\texttt{MSG2}$, which propagates $\delta_i^{k,m}$ upstream based on \eqref{delta_recursive}. 
However, \eqref{delta_no_cache} depends on $\partial J/\partial F_{ij}^{\text{o}}$ that cannot be obtained locally. We thus use a new pre-stage message \texttt{MSG1}, which propagates downstream to compute $\partial J/\partial F_{ij}^{\text{o}}$ before initiating \texttt{MSG2}.
This two-stage messaging enables fully decentralized gradient estimation even with mobility-induced tunneling. 

To estimate gradients, node $i$ obtain $d_{ij}$, $d^\prime_{ij}$, $D^\prime_{ij}$, $q_{ij}$, $\lambda_{ij}$, $r_{i}^k$ locally, and estimates $D^{\text{o}}_{i,k,m}$ via request RTT.
Then, it calculates $\tau_{i}^{k,m}$ (by \eqref{tau}), and $B_{ij}$, $m_{i}^{k,m}$ defined as:
\begin{gather}
     B_{ij} := \Lambda_i q_{ij} d_{ij}^\prime \left(\sum\nolimits_{(k,m) \in \mathcal{S}}r_{i}^{k,m}\phi_{ij}^{k,m}e^{-\Lambda_i D^{\text{o}}_{i,k,m}}\right),
    \label{Bij}
    \\  m_{i}^{k,m} := \Lambda_i r_{i}^{k,m} e^{-\Lambda_i D^{\text{o}}_{i,k,m}}\left(\sum\nolimits_{j\in\mathcal{N}_i} D^\prime_{ij}q_{ij}\right).
    \label{Mikm}
\end{gather}

Messages $\texttt{MSG1}$ are propagated downstream to calculate:
\begin{equation}
    M_i^{k,m} = \sum\nolimits_{ l \in \mathcal{N}_i} \phi_{li}^{k,m} M_{l}^{k,m} + m_i^{k,m}.
    \label{M_recursive}
\end{equation}
After obtaining $M_i^{k,m}$ for all services, node $i$ calculates
\begin{equation}
    \frac{\partial J}{\partial F_{ij}^{\text{o}}} = D^\prime_{ij} + \sum\nolimits_{k,m}L_{k,m}^{\text{res}}\phi_{ij}^{k,m}M_i^{k,m}{d^\prime_{ij}}/{(1-B_{ij})}
    \label{pJpFo}
\end{equation}

\begin{theo}
Suppose $\boldsymbol{\phi}$ is loop-free, then $\partial J/\partial F_{ij}^{\text{o}}$ is given by \eqref{pJpFo} with variable $M_i^{k,m}$ recursively defined in \eqref{M_recursive}.
\label{thm_DMP_correctness}
\end{theo}
\showiffull{
\begin{proof}
    See Appendix \ref{proof_thm_DMP_correctness}.
\end{proof}
}

\vspace{-0.3\baselineskip}

\label{subsec:localFW}
After obtaining gradients via DMP, each node independently updates its variables $(\boldsymbol{s}, \boldsymbol{\phi})$ through a Frank-Wolfe update similar to \cite{zhang2024congestion}.
We assume the network starts at $\boldsymbol{s}(0),\boldsymbol{\phi}(0)$, with loop-free $\boldsymbol{\phi}(0)$ and finite $J(0)$.
Let $\boldsymbol{s}_{i}^k = [s_{i}^{k,m}]_{m\in\mathcal{M}_k}$ and $\boldsymbol{\phi}_{i}^{k,m} = [\phi_{ij}^{k,m}]_{j \in \mathcal{V}}$, for $n \geq 0$,
\begin{small}
\begin{subequations}
\begin{align}
    \boldsymbol{s}_{i}^k(n+1) &= \boldsymbol{s}_{i}^k(n) + \alpha(n)\left(\boldsymbol{d}^{s}_{i,k}(n) - \boldsymbol{s}_{i}^k(n)\right), \label{update_FW_s}
    \\\boldsymbol{\phi}_{i}^{k,m}(n+1) &= \boldsymbol{\phi}_{i}^{k,m}(n) + \alpha(n)\left(\boldsymbol{d}^{\phi}_{i,k,m}(n) - \boldsymbol{\phi}_{i}^{k,m}(n)\right),\label{update_FW_phi}
\end{align}
\label{update_FW}
\end{subequations}
\end{small}
the update directions $\boldsymbol{d}^{s}_{i,k}(n)$ and $\boldsymbol{d}^{\phi}_{i,k,m}(n)$ are given by:
\begin{small}
\begin{equation}
\begin{aligned}
    \boldsymbol{d}^{s}_{i,k}(n) &= \arg\min\nolimits_{\boldsymbol{d}\in\mathcal{D}^s_{i,k}}\left\langle\nabla^s_{i,k} J(n),\boldsymbol{d}\right\rangle, \\
    \boldsymbol{d}^{\phi}_{i,k,m}(n) &= \arg\min\nolimits_{\boldsymbol{d}\in\mathcal{D}^{\phi}_{i,k,m}}\left\langle\nabla^{\phi}_{i,k,m}J(n),\boldsymbol{d}\right\rangle,
\end{aligned}
\label{update_FW_direction}
\end{equation}
\end{small}
$\alpha(n)$ is the step sizes; 
 $\nabla^s_{i,k}J = [\partial J/\partial s_i^{k,m}]_{m\in\mathcal{M}_k}$ and $\nabla^{\phi}_{i,k,m}J = [\partial J/\partial \phi_{ij}^{k,m}]_{j\in\mathcal{V}}$ are gradients;
 $\mathcal{D}^s_{i,k}$ and $\mathcal{D}^{\phi}_{i,k,m}$ are feasible sets, defined by
 \begin{small} 
 \begin{equation*}
 \begin{gathered}
     \mathcal{D}^s_{i,k} = \left\{\boldsymbol{s}_{i}^k \geq \boldsymbol{0}: \text{\eqref{constraint_s} holds}\right\},  
     \\ \mathcal{D}^{\phi}_{i,k,m} =\left\{\boldsymbol{\phi}_{i}^{k,m} \geq \boldsymbol{0}: \text{\eqref{constraint_phi} holds}; \, \phi_{ij}^{k,m} = 0, \forall j \in \mathcal{B}_i^{k,m}\right\}.
 \end{gathered}     
 \end{equation*}
 \end{small}
Set $\mathcal{B}_i^{k,m} \subseteq \mathcal{V}$ is the \emph{blocked node set} invented by \cite{gallager1977minimum} that guarantees $\boldsymbol{\phi}(n)$ are loop-free throughout the algorithm.
Since $\mathcal{D}^s_{i,k}$ and $\mathcal{D}^{\phi}_{i,k,m}$ are standard simplices, linear programming \eqref{update_FW_direction} admits closed-form solutions $\boldsymbol{d}^{s}_{i,k}(n) = \boldsymbol{e}_{m_{i,k}^*(n)}$ and $\boldsymbol{d}^{\phi}_{i,k,m}(n) = \boldsymbol{e}_{j_{i,k,m}^*(n)}$, where $\boldsymbol{e}_k$ is the standard basis vector with the $k$-th entry being $1$ and all others being $0$, and
\begin{subequations}
\begin{align}
    m_{i,k}^*(n) &= \arg\min\nolimits_{m' \in \mathcal{M}_k} {\partial J}/{\partial s_{i}^{k,m'}}(n),
\label{direction_FW_simplified_s}
    \\ j_{i,k,m}^*(n) &= \arg\min\nolimits_{j' \in \mathcal{Q}_i^{k,m}} {\partial J}/{\partial \phi_{ij'}^{k,m}}(n).
\label{direction_FW_simplified_phi}
\end{align}
\label{direction_FW_simplified}
\end{subequations}
This aligns with our aforementioned intuitions of \eqref{KKT_no_cache}, i.e., choosing the service/forwarding decisions with minimum marginal costs. 
Algorithm \ref{alg_update} summarizes our local update.

\vspace{-0.5\baselineskip}
\begin{algorithm}[!]
\DontPrintSemicolon
\caption{Local Frank-Wolfe update (LFW)}
\label{alg_update}
\SetKwInOut{KwOut}{Output}
\KwOut{Variables $\boldsymbol{s}(n)$, $\boldsymbol{\phi}(n)$ for $n \ge 1$.}
Determine sets $\mathcal{B}_i^{k,m}$ by network routing protocol.\;
\SetKwFor{DoAt}{At end of $n$-th slot, node $i$}{ do}{}
\DoAt{}{%
  Obtain $\partial J/\partial \boldsymbol{s}(n)$ and $\partial J/\partial \boldsymbol{\phi}(n)$ by DMP.\;
  Determine indices $m_{i,k}^*$ and $j_{i,k,m}^*$ by \eqref{direction_FW_simplified}.\;
  Update $\boldsymbol{s}_{i}^{k}(n{+}1)$ and $\boldsymbol{\phi}_{i}^{k,m}(n{+}1)$ by \eqref{update_FW}.\;
}
\end{algorithm}
\vspace{-0.5\baselineskip}

\begin{theo}
    Suppose $\nabla J$ is L-continuous, $\alpha(n)$ satisfies $\sum_{n=1}^{\infty} \alpha(n) = \infty$ and $\sum_{n=1}^{\infty} \alpha(n)^2 < \infty$, Algorithm \ref{alg_update} converges to a limit point $(\boldsymbol{s}^*,\boldsymbol{\phi}^*) = \lim_{n \to\infty}(\boldsymbol{s}(n),\boldsymbol{\phi}(n))$, where $(\boldsymbol{s}^*,\boldsymbol{\phi}^*)$ satisfies condition \eqref{KKT_no_cache}.
\label{thm_convergence}
\end{theo}
\showiffull{
\begin{proof}
    See Appendix \ref{proof_thm_convergence}.
\end{proof}
}

\section{Optimized AI Service Placement}
\label{sec:joint_placement}
\noindent\textbf{Extended system model.} We extend our framework to jointly optimize AI service placement (i.e., determining sets $\mathcal{X}_{k,m}$) via binary variable $x_i^{k,m}$, namely, $x_i^{k,m} = 1$ if $i$ hosts model $m$ for task $k$, and $0$ otherwise.
Let $\boldsymbol{x} = [x_i^{k,m}]$ denote the \emph{global service placement decision}, with the following hold
\begin{equation}    
\sum\nolimits_{(k,m)\in\mathcal{S}} L_{k,m}^{\text{mod}}\, x_i^{k,m} \leq R_i, \quad \forall\, i \in \mathcal{V},
\label{cache_cap}
\end{equation}
where $L_{k,m}^{\text{mod}}$ is the resource occupancy of $(k,m)$ and $R_i$ is the capacity of node $i$.
Then \eqref{constraint_phi} becomes
\begin{equation}
    \sum\nolimits_{j \in \mathcal{N}_i} \phi_{ij}^{k,m} = 1 - x_i^{k,m},
    \label{flow_conservation_x}
\end{equation} 
implying node $i$ either hosts service $(k,m)$ or forwards all arriving $(k,m)$ requests to its neighbors.
To address the combinatorial difficulty, we adopt a relaxation approach similar to \cite{zhang2024congestion}.
We treat $x_i^{k,m}$ as independent Bernoulli random variables with $y_i^{k,m} = \mathbb{E}[x_i^{k,m}] \in [0,1]$ being the probability that $i$ hosts $(k,m)$. We thus optimize over vector $\boldsymbol{y} = [y_i^{k,m}]$, with
\begin{equation}
    \sum\nolimits_{(k,m)} L_{k,m}^{\text{mod}} y_i^{k,m} \leq R_i, \quad y_i^{k,m}+  \sum\nolimits_{j \in \mathcal{N}_i} \phi_{ij}^{k,m} = 1.
    \label{cache_constraint_coupled}
\end{equation}
The expected computation workload at node $i$ is now
\begin{equation}
    G_i = \sum\nolimits_{(k,m)\in\mathcal{S}} W_{k,m}\,y_i^{k,m}\,t_i^{k,m}, \quad \forall i \in \mathcal{V}.
\end{equation}

With these extensions, the joint optimization problem for service placement, selection, and routing is cast as
\begin{align}
\min_{\boldsymbol{s},\boldsymbol{\phi}, \boldsymbol{y}} \quad &J = \sum_{i,j} D_{ij} + \sum_{i} C_{i} - \sum_i\sum_{k,m}\hat{u}_{k,m} r_i^k s_i^{k,m} \nonumber
 \\ \text{s.t.} \quad &  \boldsymbol{s} \geq \boldsymbol{0}, \boldsymbol{\phi} \geq \boldsymbol{0}, \boldsymbol{1} \geq \boldsymbol{y} \geq \boldsymbol{0}, \text{\eqref{constraint_s},\eqref{constraint_phi},\eqref{cache_constraint_coupled} hold}
 \tag{P2}
\label{obj_cache}
\end{align}

\begin{theo}[KKT with service placement] Suppose $(\boldsymbol{s},\boldsymbol{\phi},\boldsymbol{y})$ optimally solves \eqref{obj_cache}, then \eqref{KKT_no_cache} holds. It also holds that
\begin{equation}
 \xi_i^{k,m} \begin{cases}
        \geq  \min_{(k^\prime,m^\prime) \in \mathcal{S} : y_i^{k^\prime,m^\prime}> 0} \xi_i^{k^\prime,m^\prime}, \, \text{ if } y_i^{k,m} = 1,
         \\ \leq  \min_{(k^\prime,m^\prime) \in \mathcal{S} : y_i^{k^\prime,m^\prime}> 0} \xi_i^{k^\prime,m^\prime}, \, \text{ if } y_i^{k,m} = 0,
    \\ =  \min_{(k^\prime,m^\prime) \in \mathcal{S} : y_i^{k^\prime,m^\prime}> 0} \xi_i^{k^\prime,m^\prime}, \, \text{ o.w. }
    \end{cases}\label{KKT_cache_3}
    \end{equation}
where $\xi_i^{k,m} = \left(\min\nolimits_{j \in \mathcal{N}_i} {\partial J}/{\partial \phi_{ij}^{k,m}}\right)/L^{\text{mod}}_{k,m}$.
\label{thm_KKT_cache}
\end{theo}
Condition \eqref{KKT_cache_3} indicates that each node $i$ prioritizes hosting services based on the marginal latency reduction per unit of hosting resource, captured by $\xi_i^{k,m}$.
Moreover, Theorem \ref{thm_decentralize} applies with the recursive decomposition \eqref{delta_recursive} generalized to
\begin{small}
\begin{equation*}
    \delta_i^{k,m} = y_i^{k,m}W_{k,m}C'_i + \sum_{j \in \mathcal{N}_i}\phi_{ij}^{k,m}\big(L_{k,m}^{\text{req}}\frac{\partial J}{\partial F_{ij}^{\text{o}}}+ L_{k,m}^{\text{res}}\frac{\partial J}{\partial F_{ji}^{\text{o}}}+\delta_{j}^{k,m}\big).
\end{equation*}
\end{small}

\begin{figure*}[]
  \begin{minipage}{.43\linewidth}
    \centering
\begin{tabularx}{\textwidth}
{c|ccccccc}
  \toprule
  Name & $|\mathcal{V}|$ & $|\mathcal{K}|$ & $|\mathcal{S}|$ & $\mu_{ij}$ & $\nu_i$ & $\Lambda_i$ & $R_i$ \\
  \midrule
  \texttt{grid}  & 9   & 5  & 15  & 10  & 10  & 0.10 & 20 \\
  \texttt{MEC}   & 13  & 5  & 20  & 10  & 10  & 0.10 & 20 \\
  \texttt{ER}    & 30  & 20 & 40  & 15  & 15  & 0.15 & 30 \\
  \texttt{D-Tel} & 68  & 30 & 100 & 15  & 15  & 0.15 & 30 \\
  \texttt{SW} & 120  & 45 & 150 & 20  & 20  & 0.15 & 30 \\
  \bottomrule
\end{tabularx}
\vspace{-0.3\baselineskip}
    \captionof{table}
      { Scenarios
        \label{tab:scenarios}
      }
  \end{minipage}\hfill
    \begin{minipage}{.57\linewidth}
    \centering
    \vspace{-0\baselineskip}
    \includegraphics[width=0.98\textwidth]{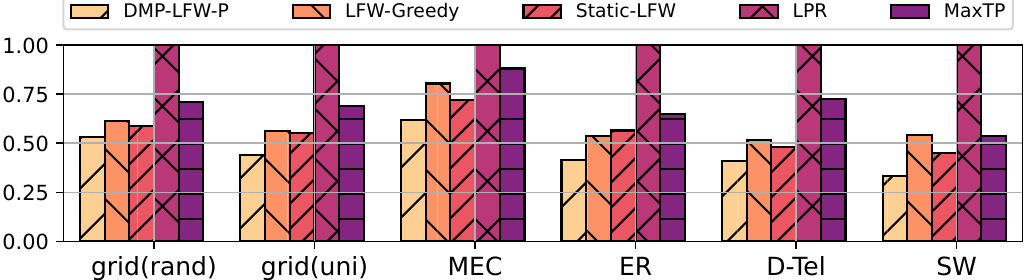}%
    \vspace{-0.1\baselineskip}
    \caption
      {%
        Normalized objective $J$ in all scenarios
        \label{fig:barfig}%
      }%
  \end{minipage}
\vspace{-1\baselineskip}
\end{figure*}

\begin{figure*}[]
     \begin{minipage}{.28\linewidth}
    \centering
    \vspace{-0\baselineskip}
    \includegraphics[width=0.98\textwidth]{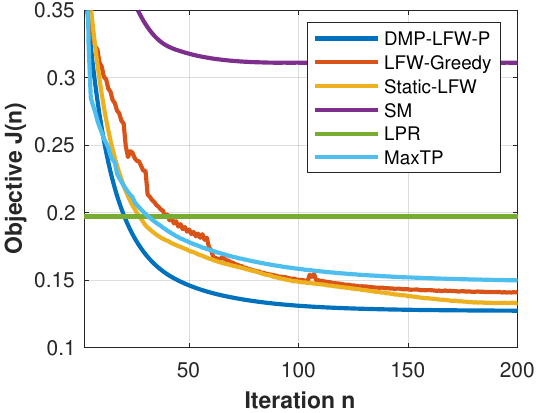}%
    \vspace{-0.3\baselineskip}
    \caption
      {%
        Convergence trajectory
        \label{fig:convergence}%
      }%
  \end{minipage}\hfill
    \begin{minipage}{.16\linewidth}
    \centering
    \vspace{-0\baselineskip}
    \includegraphics[width=0.98\textwidth]{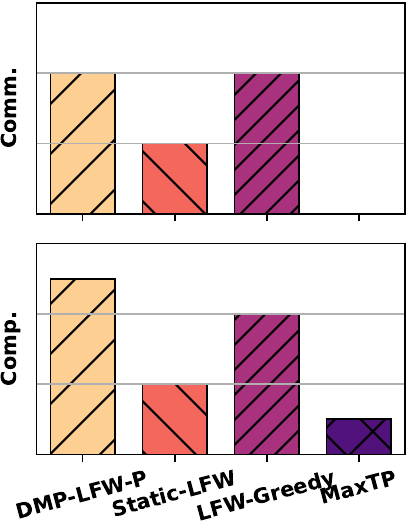}%
    \vspace{-0.1\baselineskip}
    \caption
      {Overhead
        \label{fig:over}%
      }%
  \end{minipage}\hfill
    \begin{minipage}{.265\linewidth}
    \centering
    \vspace{-0\baselineskip}
    \includegraphics[width=0.98\textwidth]{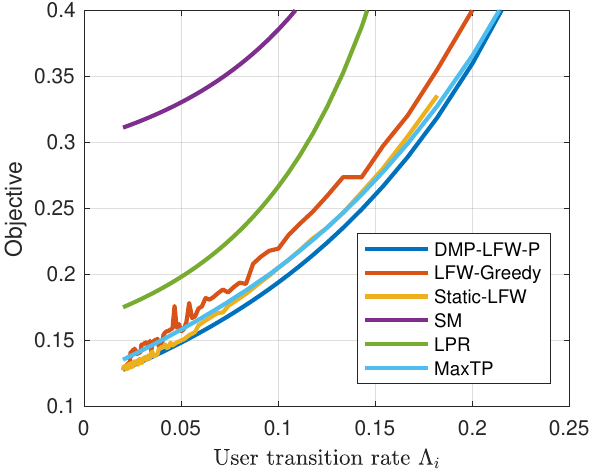}%
    \vspace{-0.3\baselineskip}
    \caption
      {%
        $J$ v.s. $\Lambda_i$
        \label{fig:mobility}%
      }%
  \end{minipage}\hfill
  \begin{minipage}{.265\linewidth}
    \centering
    \vspace{-0\baselineskip}
    \includegraphics[width=0.98\textwidth]{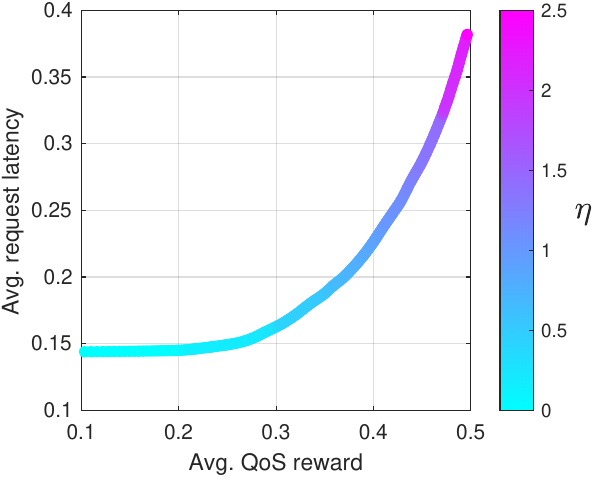}%
    \vspace{-0.3\baselineskip}
    \caption
      {%
       QoS-latency tradeoff
       \label{fig:tradeoff}%
      }%
  \end{minipage}
\vspace{-1\baselineskip}
\end{figure*}

We omit the decentralized algorithm for the optimized AI service placement case as it is almost a replica of Sec \ref{sec:decentralized_alg}.
With $\boldsymbol{y}$ involved, the simplification \eqref{direction_FW_simplified_phi} no longer holds. We present in \cite{supplementary2025} a valid simplification for Frank-Wolfe update.

\section{Numerical Evaluation}
\label{sec:simulation}

{
We conduct a flow-level numerical evaluation of the proposed algorithm and baselines on both synthetic and real-world network topologies:
\textbf{\texttt{grid}}: A $5 \times 5$ grid network.
\textbf{\texttt{MEC}}: A 3-level 3-ary tree with same-parent nodes linearly connected, representing a typical hierarchical MEC architecture.
\textbf{\texttt{ER}}: A connectivity-guaranteed Erdős-Rényi graph with edge probability $p = 0.15$.
\textbf{\texttt{D-Tel}}: The backbone topology of Deutsche Telekom.
\textbf{\texttt{SW}}: A Watts-Strogatz small-world network.

We summarize key parameters of these scenarios in Table~\ref{tab:scenarios}. Moreover, we assume $r_i^k = 1$, $L^{\text{req}}_{k,m} = 0.25$, $L^{\text{res}}_{k,m} = 0.75$, and $L^{\text{mod}}_{k,m}$ follow the sequence $[10, 20, 30, \ldots]$, with corresponding utilities $u_{k,m} = [0.1, 0.3, 0.5, \ldots]$. Delays on links and nodes are approximated via a third-order Taylor expansion, with $\mu_{ij}$ and $\nu_i$ as service rates. Mobility transition probabilities $q_{ij} \in [0,1]$ are u.a.r. with $\sum_j q_{ij} = 1$. We set the latency–utility tradeoff parameter to $\eta = 1$.

Beyond the proposed decentralized protocol \textbf{\texttt{DMP-LFW-P}}, we implement the following baselines:
\textbf{\texttt{LFW-Greedy}}: Uses \texttt{DMP} and \texttt{LFW}, with each node greedily serving the most popular services (based on $t_i^{k,m}$) until capacity is filled.
\textbf{\texttt{Static-LFW}}: A static variant of \cite{zhang2024congestion}, approximating gradients as $\partial J / \partial F^{\text{o}}_{ij} = D'_{ij}$ without propagating \texttt{MSG1}, thus ignoring tunneling.
\textbf{\texttt{SM}}: Models latency under service migration, assuming entire models are transferred between same-layer nodes upon user transitions.
\textbf{\texttt{LPR}}~\cite{liu2020distributed}: Solves a linear program for model selection and routing under greedy placement, using marginal delays $d_{ij}|_{F_{ij} = 0}$ and $c_i|_{G_i = 0}$.
\textbf{\texttt{MaxTP}}: A flow-level approximation of backpressure-based scheduling that minimizes the maximum local queue size.

We set $\alpha = 0.05$ and construct $\mathcal{B}_i^{k,m}$ to resemble a service-specific DAG with maximal edge coverage. Figure~\ref{fig:barfig} shows the normalized convergent $J$ across scenarios excluding \texttt{SM}. For \texttt{grid}, we compare \texttt{grid(rand)} (random $q_{ij}$) and \texttt{grid(uni)} (uniform $q_{ij}$). \texttt{DMP-LFW-P} consistently outperforms all baselines, achieving up to 17\% improvement over the second best. Gains are more prominent under directional user movement. \texttt{LFW-Greedy} and \texttt{Static-LFW} ablate joint service placement and tunneling awareness. \texttt{LPR} performs the worst by ignoring congestion, while \texttt{MaxTP} is second worst due to not directly optimizing latency. Notably, our method yields increasing benefits as network scale grows.

We further investigate \texttt{grid} in detail. Figure~\ref{fig:convergence} illustrates convergence trajectories. Figure~\ref{fig:over} compares per-node communication and computation overheads. All distributed methods exhibit per-node complexity $O(|\mathcal{S}||\mathcal{N}_i|)$; we evaluate computational load via its coefficient and communication via average control messages exchanged.
Figure~\ref{fig:mobility} shows objective $J$ versus user transition rate $\Lambda_i$. As mobility increases (e.g., $\Lambda_i \geq 0.1$), congestion sharply rises, degrading performance across all methods. In this high-mobility regime, \texttt{MaxTP} approaches the performance of \texttt{DMP-LFW-P}.

Figure~\ref{fig:tradeoff} illustrates the tradeoff between QoS and latency under different preference $\eta$. Each point on the curve reflects the converged state of \texttt{DMP-LFW-P}. Higher $\eta$ leads to increased average QoS, but also to superlinearly growing latency, indicating an increasing marginal delay for each QoS gain.

}

\bibliographystyle{IEEEtran}
\bibliography{References}

\end{document}